\documentclass[aps,11pt,nofootinbib,preprintnumbers,showpacs,superscriptaddress]{revtex4}

 \usepackage{here}

\usepackage[dvipdfmx]{graphicx}
\usepackage{amssymb,amsfonts,amsmath,bm,color,multirow,cases,empheq,hyperref}
\usepackage{mathrsfs}

\usepackage{enumerate}
\usepackage{ulem}
\usepackage{afterpage}
\usepackage{ulem}
\usepackage{wrapfig}
\usepackage{ulem}

\newcommand{\del}[1]{ \partial_{#1} }

\hypersetup{%
 setpagesize=false,
 bookmarksnumbered=true,%
 bookmarksopen=true,%
 colorlinks=true,%
 linkcolor=blue,%
 citecolor=red}

\begin{document}

\title{\Large 
 Conversion of gravitational and electromagnetic waves without any external background fields
}

\author{Takashi Mishima${}^1$\footnote{mishima.takashi@nihon-u.ac.jp} and Shinya Tomizawa${}^{2}$\footnote{tomizawa@toyota-ti.ac.jp}}
\vspace{2cm}
\affiliation{
${}^1$Laboratory of Physics, College of Science and Technology, Nihon University,
Narashinodai, Funabashi, Chiba 274-8501, Japan\\
${}^2$Mathematical Physics Laboratory, Toyota Technological Institute,  Hisakata 2-12-1, Nagoya 468-8511, Japan
}
\preprint{TTI-MATHPHYS-10}

\begin{abstract}
Applying a simple harmonic map method to the cylindrically symmetric Einstein-Maxwell system, we obtain exact solutions representing strong nonlinear interaction between gravitational waves and electromagnetic waves in the case without any background field. 
As an interesting fact, we can show that with adjusted parameters the solutions represent occurrences of large conversion phenomena in the intense region of fields near the cylindrically symmetric axis.

\end{abstract}

\maketitle

The first  direct confirmation of the existence of gravitational waves \cite{LIGOScientific:2016aoc} has inspired a lot of  fundamental study of gravitational waves from various perspective. 
As the observation accuracy improves in the future, it will become more and more important to know the fundamental aspects of gravitational waves  themselves. 
In particular, clarification of the nonlinear behavior of gravitational waves in the regions with intense fields will continue to be one of the important clues to explore new gravitational physics. 
For this purpose, we have robust numerical techniques which have already being used vigorously. 
On the one hand, we have a long history of studying non-linear aspects of gravity using exact solutions and have an accumulated heritage~\cite{bicak:2000, book exact solution, book Exact Space-Times, book colliding waves}. 
The latter way will be still useful to add new insights matching this new trend if the appropriate ingenuity is introduced.
From a more realistic point of view, the study of gravitational waves combined with other fields will be needed more in the future. 
Because of the universality of the gravitational interaction with matters, 
it may be natural to expect that as nonlinear phenomena, various conversions between gravitational waves and matter fields occur in regions with intense fields. 
Hence, as a simple and important example, we consider a Einstein-Maxwell system, that is, a coupled system with gravitational and electromagnetic waves. 
In fact, by using linear perturbation methods, so far several studies have been done to treat the occurrence of conversion phenomena based on the existence of external background fields, such as electromagnetic fields around black holes~\cite{Gerlach:1974zz, Olson:1974nk, Matzner:1976kj, Crispino:2009zza, Saito:2021sgq, Hadj:2021yry}.
On the other hand, within the scope of linear perturbative method, such conversions cannot be expected in the limit of neutral vacuums, because the mixing tendency of the waves corresponding to the perturbed Einstein-Maxwell system becomes weak and eventually disappears in such cases~\cite{Gerlach:1974zz, Olson:1974nk, Matzner:1976kj, Crispino:2009zza, Zerilli:1974ai, Moncrief:1974gw, Moncrief:1974ng, Moncrief:1975sb}. 
From this, as a guess, the conversions from gravitational fields to electromagnetic fields seem to be not so easy phenomena to occur in comparison with the reverse conversions.
So we take interest in how and to what extent the conversion occurs between the gravitational wave and the electromagnetic wave even without any background field by a fully non-linear analysis based on the exact solutions.
It should be noticed that possibility and importance to analyze the conversion phenomena by using the exact solutions have been already pointed out by Alekseev~\cite{Alekseev:2015wbn}, and also the possibility of using cylindrical symmetric solutions has been suggested by us~\cite{Mishima:2017zpn}. 
Our strategy is the following: we first construct simple exact solutions which represent the waves corresponding to the Einstein-Maxwell system under cylindrical symmetry by using a harmonic method; then use the solutions to reveal the novel and interesting aspects of full nonlinearity of the Einstein-Maxwell system.
Especially, here we concentrate on some features of the conversion phenomena, as mentioned above. 

\medskip
We assume the following useful line element~\cite{Kompaneets-Jordan-Ehlers} (see \cite{Bronnikov:2019clf} for general treatments of  cylindrical symmetric systems) and gauge potential ${\bf A}$:
\begin{eqnarray}
ds^2
 &=& e^{2\psi}(dz-w d\phi)^2+\rho^2 e^{-2\psi}d\phi^2+e^{2(\gamma-\psi)}(-dt^2+d\rho^2),  
  \label{eq:KJE}  \\
{\bf A}
 &=& A_{\phi}d\phi + A_{z}dz.  
\end{eqnarray}
The metric functions $\psi$, $w$, $\gamma$ and gauge potentials $ A_{\phi}$, $ A_{z}$ depend only on the coordinates $t$ and $\rho$. 
Here, 
we also introduce a twist potential $\Phi$ and a `magnetic' potential $\chi$ corresponding to $w$ and $ A_{\phi}$, respectively. 
These are defined by $d\chi :=-i_{\xi}\ast {\bf F} $ and $d\Phi := \ast (\xi\land d\xi)-2(A_{z}d\chi-\chi dA_{z})$, where $\xi$ corresponds to a Killing vector $\del{}/\del{}z$ and ${\bf F}$ is an electromagnetic field.
Then the cylindrically symmetric Einstein-Maxwell equations can be reduced to the so-called Ernst equation~\cite{Ernst:1967by} by introducing the following two complex potentials $E$ and $F$, 
\begin{eqnarray}
E:= e^{2\psi} +|F|^2- i \Phi,\ \ \ F:= A_{z} + i\chi .
  \label{eq:Ernst pot}
\end{eqnarray}
For convenience, we however adopt another type of the Ernst equation in the following~\cite{Ernst:1967by, Kinnersley73}: 
\begin{eqnarray}
(\xi \bar{\xi}+\eta \bar{\eta}-1 )\, \nabla^2{\xi}
&=& 2(\bar{\xi} \nabla{\xi} + \bar{\eta} \nabla{\eta} )\cdot \nabla{\xi}, 
  \label{eq:Ernst1} \\
(\xi \bar{\xi}+\eta \bar{\eta}-1 )\, \nabla^2{\eta}
&=& 2(\bar{\xi} \nabla{\xi} + \bar{\eta} \nabla{\eta} )\cdot \nabla{\eta},
  \label{eq:Ernst2}
\end{eqnarray}
where the complex potentials $\xi$ and $\eta$ are related to the potentials $E$ and $F$, as follows (see, for example, \cite{ book colliding waves}) 
\begin{eqnarray}
\xi= \frac{E-1}{E+1},\ \ \ \ \eta= \frac{2F}{E+1},
  \label{eq:Relation1}
\end{eqnarray}
and $\nabla$ is the gradient defined on the three dimensional Minkowski space $M^{(1,2)}$.

Once the solutions of Eqs.~(\ref{eq:Ernst1}) and~(\ref{eq:Ernst2}) are given, 
through the above relations~(\ref{eq:Ernst pot}) and~(\ref{eq:Relation1}), the quantities $e^{2\psi}$, $\Phi$, ${\chi}$ and $A_{z}$ can be derived algebraically.
Then, the metric components $w$, $\gamma$ and gauge field component $A_{\phi}$ are obtained after some integral calculation.

To construct a new solution, it is convenient to regard the basic equations~(\ref{eq:Ernst1}) and~(\ref{eq:Ernst2}) as a harmonic map equation (more precisely, a wave map equation). 
So the solutions can be realized as harmonic maps from the base space $M^{(1,2)}$ to the final target space which corresponds to the ball model of complex two dimensional hyperbolic space ${\rm H^{2}_{C}}$ (hereafter ${\rm H^{2}_{C}}$ means the ball model) \cite{book Complex Hyperbolic Geometry, Parker2010}.
The field variables $\xi$ and $\eta$ of Eqs.~(\ref{eq:Ernst1}) and~(\ref{eq:Ernst2}) correspond to the two complex coordinates used to describe ${\rm H^{2}_{C}}$.
As a simple method, we adopt here the composite harmonic method~\cite{EellsSampson64}.
In short, the procedure of the method is the following: first construct a harmonic map transforming the base space to an intermediate target space (the Poincare disc model of complex one dimensional hyperbolic space ${\rm H^{1}_{C}}$ adopted here), next find out an appropriate totally geodesic embedding map of ${\rm H^{1}_{C}}$ into the ${\rm H^{2}_{C}}$, and finally combine these maps. 

As a first step, we adopt the following simple solution given in the previous work~\cite{Mishima:2017zpn}: 
\begin{eqnarray}
\xi_{v} = \frac{1-e^{-2\tau}+iA}{1+e^{-2\tau}-iA}.  
  \label{eq:MT17-1}
\end{eqnarray}
Here the constant $A$ is real and the subscript $v$ means a solution of the vacuum Ernst equation (
i.e. Eq.~(\ref{eq:Ernst1}) for $\eta=0$). 
The solution $\xi_{v}$ corresponds to the complex coordinate of the ${\rm H^{1}_{C}}$. 
And also the real parameter $\tau$ is replaced with a cylindrically symmetric wave function which satisfies the linear wave equation defined on the three dimensional Minkowski space $M^{(1,2)}$, so that the wave function $\tau$ can be considered as a seed function. 
As a simple totally geodesic embeddings of ${\rm H^{1}_{C}}$, we use the subspace of ${\rm H^{2}_{C}}$ defined by $(\xi, \eta) = (\cos2\theta\, z,\sin2\theta\,  z)$,  where the parameter $z$ is considered as a complex coordinate of the ${\rm H^{1}_{C}}$.
In fact, the similar ansaz has been already used by Halilsoy for a different situation to treat colliding plane waves~\cite{Halilsoy:1989ji} (see also \cite{book colliding waves}).
Then we obtain the harmonic map by replacing the parameter $z$ of the subspace introduced  above with the function $\xi_{v}$ defined by  Eq.~(\ref{eq:MT17-1}), 
as follows 
\begin{eqnarray}
 (\xi, \eta) = (\cos2\theta\, \xi_{v},\sin2\theta\,  \xi_{v}). 
   \label{eq:TGEmbd}
\end{eqnarray}
From this harmonic map, the expression of the Ernst potentials is given after some algebraic calculation as below: 
\begin{eqnarray}
e^{2\psi}
 &=& \frac{1}
       {A^2 e^{2\tau} \cos^4 \theta + e^{-2\tau} (\cos^2 \theta +e^{2\tau} \sin^2 \theta)^2 }, \nonumber   \\
\Phi
 &=& -{A e^{2\tau}\cos 2\theta }\,e^{2\psi},   \ \ \ 
\chi
 =  \frac{1}{2} {A e^{2\tau}\sin 2\theta }\, e^{2\psi},   \nonumber \\
A_{z}
 &=& -\frac{1}{2} 
   { [ A^2 e^{2\tau}\cos^2 \theta + ( e^{-2\tau}-1)( \cos^2 \theta + e^{2\tau}\sin^2 \theta) ] \sin 2\theta}\,e^{2\psi} . 
     \label{eq:sol-1} 
\end{eqnarray}

To clarify the nonlinear aspects using these expressions, let us consider the metric function $\gamma$ as the C-energy~\cite{Thorn65}, which is extended to cylindrically symmetric Einstein-Maxwell system.
The corresponding C-energy density ${\cal E}$ is given by the derivative of the function $\gamma$ with respect to $\rho$. The useful formula is deduced as below from the Einstein equation: 
\begin{eqnarray}
{\cal E}_{}&:=& \del{\rho}\gamma
= \frac{\rho}{8} 
\left( {\cal A}_{+}^2 + B_{+}^2 + {\cal A}_{\times}^2 + B_{\times}^2 
+ {\cal A}_{z}^2 + {\cal B}_{z}^2 + {\cal A}_{\phi}^2 + {\cal B}_{\phi}^2 \right).
\label{eq:Cenergy-1} 
\end{eqnarray}
Here, according to the previous work by Piran, Safier, and Stark~\cite{Piran:1985dk}, the following quantities, called `amplitudes', are introduced with an extension to the Einstein-Maxwell system:
\begin{eqnarray}
{\cal A}_{+} &:=& 2\ \del{v}\psi ,\ \ 
{\cal A}_{\times} := \frac{1}{\rho}e^{2\psi}\del{v}w= e^{-2\psi} \bigl[ \del{v}\Phi +2(A_{z}\del{v}\chi - \chi\del{v}A_{z} ) \bigr],    \\
{\cal A}_{z} &:=& 2e^{-\psi} \del{v}A_{z}, \ \ \ 
{\cal A}_{\phi} := \frac{2}{\rho}e^{\psi}\left(\del{v}A_{\phi}+w\del{v}A_{\phi} \right) = 2e^{-\psi} \del{v}\chi,    \\
{\cal B}_{+} &:=& 2\ \del{u}\psi , \ \ 
{\cal B}_{\times} := \frac{1}{\rho}e^{2\psi}\del{u}w = e^{-2\psi} \bigl[ \del{u}\Phi +2(A_{z}\del{u}\chi - \chi\del{u}A_{z} ) \bigr] ,    \\
{\cal B}_{z} &:=& 2e^{-\psi} \del{u}A_{z},\ \ \ 
{\cal B}_{\phi}:= \frac{2}{\rho}e^{\psi}\left(\del{u}A_{\phi}+w\del{u}A_{\phi} \right) = 2e^{-\psi} \del{u}\chi, 
\end{eqnarray}
where the symbols ${\cal A}_{(\cdot)}$ and ${\cal B}_{(\cdot)}$ mean `ingoing' and `outgoing', respectively, and the subscripts indicate which mode the amplitude corresponds to, and also the null coordinates $u=(t-\rho)/2$ and $v=(t+\rho)/2$ are introduced. 
Let us divide the C-energy density ${\cal E}$ given by Eq.~(\ref{eq:Cenergy-1}) into the two parts, the first four terms and its rest, which are referred to as ${\cal E}_{\rm g}$ and ${\cal E}_{\rm em}$ in the following, respectively.
${\cal E}_{\rm g}$ can be considered a gravitational part, according to the work~\cite{Piran:1985dk}. On the other hand ${\cal E}_{\rm em}$ can be regarded as an electromagnetic part, since ${\cal E}_{\rm em}$ actually corresponds to the $tt$-component of electromagnetic energy momentum tensor.
As a next step, let us introduce 
the occupancy ratios $R_{\rm g}$ and $R_{\rm em}$ ($R_{\rm g}+R_{\rm em}=1$), which correspond to the gravitational and electromagnetic parts, respectively, as follows,
\begin{eqnarray}
{\cal E}_{\rm g} &=& R_{\rm g}{\cal E},\ \ \ 
R_{\rm g}:= 
\frac{ A^2e^{4\tau}\cos^4\theta + ( \cos^2\theta - e^{2\tau}\sin^2\theta )^2 }
       { A^2e^{4\tau}\cos^4\theta + ( \cos^2\theta + e^{2\tau}\sin^2\theta )^2 }, 
        \label{eq:Cenergy-g}      \\
{\cal E}_{\rm em} &=& R_{\rm em}{\cal E},\ \ \ 
R_{\rm em}:= 
\frac{  e^{2\tau}\sin^2 (2\theta) }
       { A^2e^{4\tau}\cos^4\theta + ( \cos^2\theta + e^{2\tau}\sin^2\theta )^2 },  
           \label{eq:Cenergy-em}\\
{\cal E}_{} &=& \rho\left[\, (\del{t}\tau)^2 + (\del{\rho}\tau)^2 \, \right],
\end{eqnarray}
where 
the explicit forms of $R_{\rm g}$, $R_{\rm em}$, and ${\cal E}$ are derived from Eq.~(\ref{eq:sol-1}). 
It should be noticed that the total C-energy density ${\cal E}$ itself 
depend on neither of $A$ nor $\theta$
because of the invariance of ${\cal E}$ under the isometry of the target space. 
So, the non-trivial effects of non-linear interaction between gravitational modes and electromagnetic modes will appear only 
 through the occupancy ratios $R_{\rm g}$ and $R_{\rm em}$.
These quantities show how the seed function $\tau$ determines the local spacetime dependence of the occupancy ratio of each contribution, once the parameters $(A, \theta)$ are fixed.
%
\begin{figure}[h]
  \begin{tabular}{ccc}
 \begin{minipage}[t]{0.40\hsize}
 \centering
\includegraphics[width=6cm]{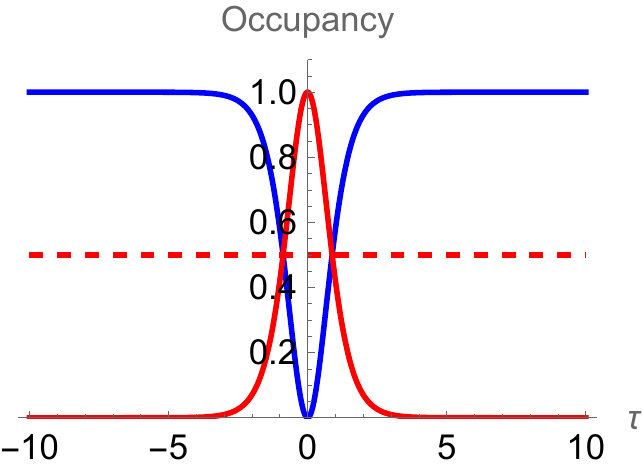}
 \end{minipage} &\ \ \ \ \ \ 
 \begin{minipage}[t]{0.40\hsize}
 \centering
\includegraphics[width=6cm]{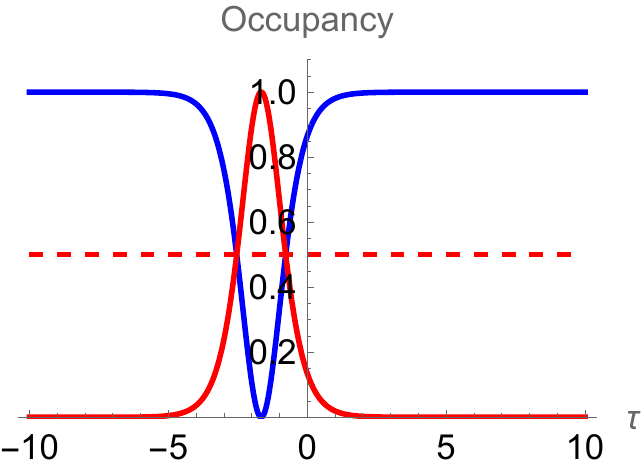}
 \end{minipage} &\ \ \ \ \ \ 
  \end{tabular}
\caption{
The left and right figures show the occupancy ratios which correspond to the cases $(A, \theta)=(0, \pi/4)$ and $(A, \theta)=(1/6, 11\pi/25)$, respectively. Blue and red graphs correspond to the gravitational and electromagnetic contributions, respectively.
The horizontal axis corresponds to the seed function $\tau$. 
For the case of asymptotically flat spacetime, the infinities correspond to $\tau=0$. 
}
\label{fig:fig1}
\end{figure}

To consider the conversion phenomena,  let us introduce {\it occupancy diagrams}
 by using the formulas~(\ref{eq:Cenergy-g}) and~(\ref{eq:Cenergy-em}), as depicted in Fig.~\ref{fig:fig1}. 
From the occupancy diagrams, we can qualitatively predict how the conversion phenomenon will occur according to the behavior of the seed function $\tau$.
Each set of the parameters $(A, \theta)$ gives one occupancy diagram, which corresponds to one of various patterns of the conversion  phenomenon. 
Especially, if the parameters $(A, \theta)$ are  appropriately adjusted, the graphs of the occupancy ratio of the electromagnetic part show a very large peak at some value of $\tau$. 
This makes us expect that the electromagnetic part of the C-energy can become overwhelmingly dominant over the gravitational part at some time. 
For example, in the graphs of Fig.~\ref{fig:fig1}, such a peak appears at $\tau=0$ for the left case, and near $\tau=-1.6$ for the right case. 

Here let us note the following fact: when a regular wave packet like the Weber-Wheeler-Bonnor solution (WWB)~\cite{Weber:1957oib,bonnor:1957} is adopted as the seed function, in general at null  infinity the seed  function becomes zero and near the cylindrical symmetric axis $\rho=0$ the seed  function can have any large values.
From this, we can immediately deduce the following consequences, for example. 
First, let us assume that the seed function is a wave packet like the WWB solution, whose peak is initially put on the symmetric axis and take a value of about $\tau=5$.
Then, from the left figure of Fig.~\ref{fig:fig1}, it is natural to expect that most of the corresponding wave is composed of gravitational parts near the axis initially, and then the wave rapidly changes to a wave composed mainly of electromagnetic parts as it spreads and decays.
Similarly, from the right figure in Fig.~\ref{fig:fig1}, under the assumption that the initial peak of the seed takes a value of about $-1.6$ on the axis, it can be expected that the corresponding initial configuration of the wave is mainly composed of electromagnetic parts, and then the wave is converted to the gravitational wave after leaving the axis.

\begin{figure}[h]
  \begin{tabular}{ccc}
 \begin{minipage}[t]{0.30\hsize}
 \centering
\includegraphics[width=5cm]{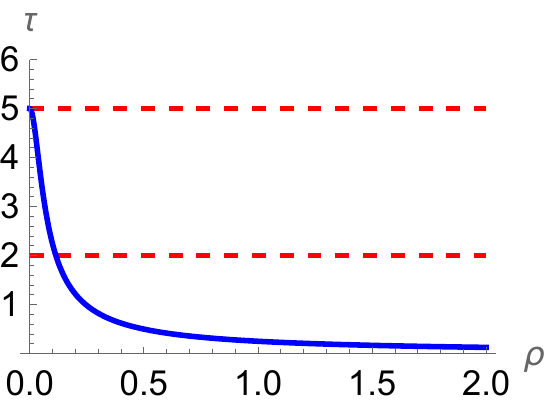}
 \end{minipage} &\ \ \ \ \ \ 
 \begin{minipage}[t]{0.30\hsize}
 \centering
\includegraphics[width=5cm]{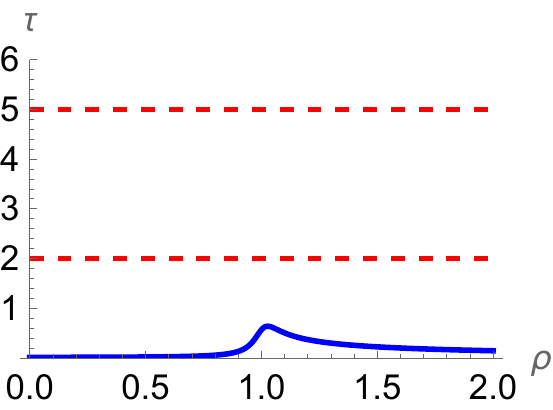}
 \end{minipage} &\ \ \ \ \ \ 
  \end{tabular}
\caption{
The left and right figures display the snapshots of the seed function $\tau$, which correspond to $t=0$ and $t=1$, respectively. The parameters ($a$, $c$) set to (1/20, 1/4). }
\label{fig:fig2}
\end{figure}

\begin{figure}[h]
  \begin{tabular}{ccc}
 \begin{minipage}[t]{0.30\hsize}
 \centering
\includegraphics[width=5cm]{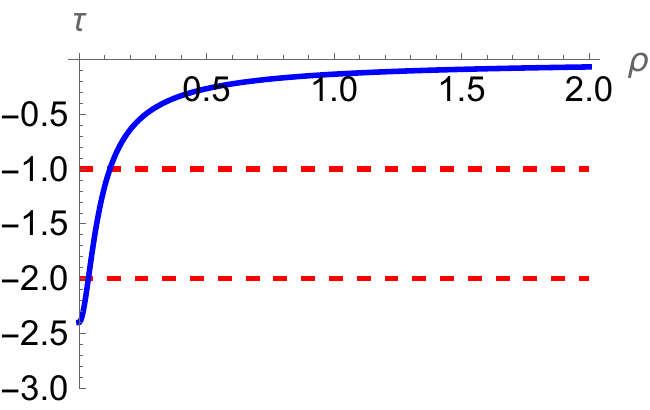}
 \end{minipage} &\ \ \ \ \ \ 
 \begin{minipage}[t]{0.30\hsize}
 \centering
\includegraphics[width=5cm]{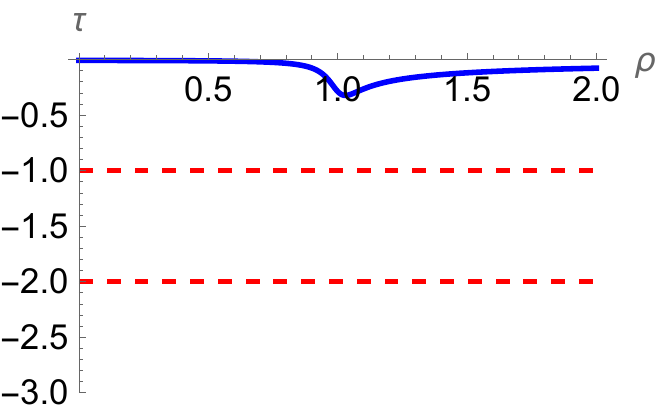}
 \end{minipage} &\ \ \ \ \ \ 
  \end{tabular}
\caption{
The left and right figures display the snapshots of the seed function $\tau$, which correspond to $t=0$ and $t=1$, respectively. The parameters ($a$, $c$) set to (1/18, -2/15). }
\label{fig:fig3}
\end{figure}

To proceed the analysis further, we should give the seed function $\tau$ explicitly.
Hence in the rest, we adopt the following expression of WWB solution as a seed function $\tau$ (for explicit forms, see \cite{Ashtekar:1996cm, Mishima:2017zpn}),
\begin{eqnarray}
\tau(t,\rho) 
&=&
\frac{c}{\sqrt{2}}
\left[
 \frac{\sqrt{4 a^2 t^2 + (a^2 + \rho^2 - t^2)^2}
      +a^2 + \rho^2 - t^2}
      {4 a^2 t^2 + (a^2 + \rho^2 - t^2)^2}
\right]^{1/2},
    \label{eq:WWBeq} 
\end{eqnarray}
two shapes of which, for examples, are given in Fig.~\ref{fig:fig2} and Fig.~\ref{fig:fig3} for two different sets of parameters $(a,c)$.
\begin{figure}[h]
  \begin{tabular}{ccc}
 \begin{minipage}[t]{0.40\hsize}
 \centering
\includegraphics[width=6cm]{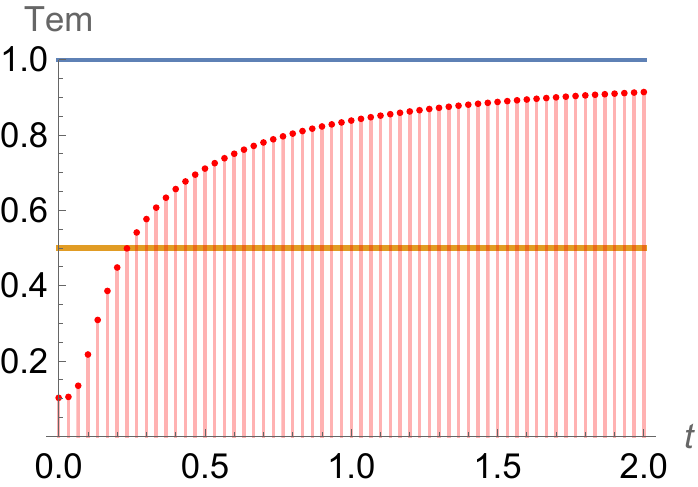}
 \end{minipage} &\ \ \ \ \ \ 
 \begin{minipage}[t]{0.40\hsize}
 \centering
\includegraphics[width=6cm]{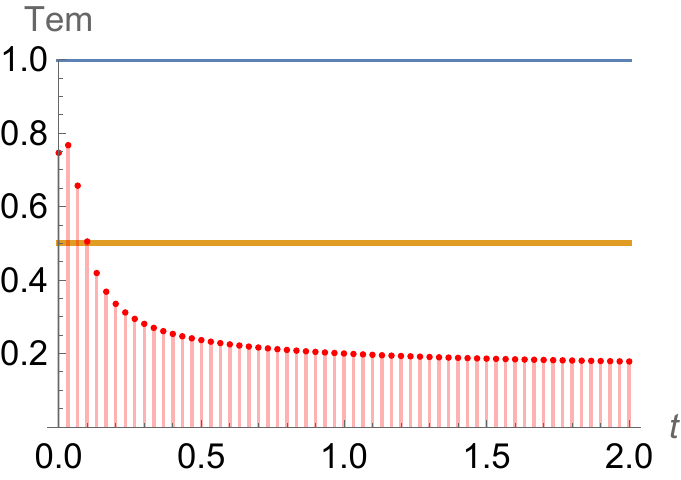}
 \end{minipage} &\ \ \ \ \ \ 
  \end{tabular}
\caption{
The left and right figures show the time dependence of $T_{\rm em}$ for the cases:  $(A, \theta, a, c)=(0, \pi/4, 1/20, 1/4)$ or $(A, \theta, a, c)=(1/6, 11\pi/25,1/18, -2/15)$. The Red line segments on both the figures mean rates of electromagnetic contributions. }
\label{fig:fig4}
\end{figure}

Using the explicit form of the seed given above, let us define the total conversion ratio $T_{\rm em}$, which is useful to show the time dependence of total electromagnetic contributions to the C-energy, as follows
\begin{eqnarray}
T_{\rm em}(t) := \frac{{\gamma}_{\rm em}(t,\rho=\infty) }{{\gamma}(t,\rho=\infty)}, 
\ \ \ \bigl(\, \gamma_{\bullet}(t, \rho) := \int_{0}^{\rho} {\cal E}_{\bullet}(t, r)\ dr \, \bigr).
    \label{eq:Tem}
\end{eqnarray}
Here the quantity ${\gamma}(t,\rho=\infty)$ takes a time-independent value $\left({c}/{2a}\right)^2$.
Figure~\ref{fig:fig4} shows, for example, the time  dependence of the $T_{\rm em}$ from $t = 0$ to $t = 2$ for $(A, \theta, a, c)=(0, \pi/4, 1/20, 1/4)$ or $(A, \theta, a, c)=(1/6, 11\pi/25,1/18, -2/15)$, respectively.
Here, to evaluate the integral numerically we have changed the upper limit of the integration range from infinity to a sufficiently large finite value (here $\rho = 1000$). 
The left figure in Fig.~\ref{fig:fig4} shows the ratio $T_{\rm em}$ increases rapidly, while the right figure shows the ratio $T_{\rm em}$ decreases promptly. 
After sufficiently long time (e.g., from $t=0$ to $t=200$),  for the left-hand case, the {\it electromagnetic} contribution is amplified times about 9.8 (i.e., ${\gamma}_{\rm em}(t=200,\rho=1000)/ {\gamma}_{\rm em}(t=0,\rho=1000)$ ), whereas, for the right-hand case, {\it the gravitational} contribution is amplified times about 3.4 (i.e., ${\gamma}_{\rm g}(t=200,\rho=1000)/ {\gamma}_{\rm g}(t=0,\rho=1000)$ ).

To summarize, 
we have clarified the occurrence of large conversion phenomena between gravitational and electromagnetic waves as a novel aspect of full nonlinear interaction of Einstein-Maxwell system.
There is already a similar attempt to treat the conversion based on the numerical analysis~\cite{Barreto:2017shm}. 
However, their attempt is interesting but somewhat limited because of the research that deals only with the conversion from electromagnetic waves to gravitational waves.
As mentioned above, one of our main concerns is about how large an initially small electromagnetic wave can be amplified being supplied with the gravitational wave energy.
In the case of the above electromagnetic wave amplification, the amplification factor is about 10 times. 
Though its factor can be considered large enough, we expect, however, that any larger amplification factor can be achieved. 
From the diagrams of Fig.~\ref{fig:fig1}, we easily know that the electromagnetic contribution decreases sharply as the magnitude of the wave function $|\tau|$ becomes larger. 
Therefore, considering that the size of the wave function $|\tau|$ is controlled by the parameter $c$ of the WWB solution (\ref{eq:WWBeq}), we may expect that the amplification factor of the electromagnetic wave will increase without limit as the size of the parameter $c$ becomes larger. 
In fact, for the case of $(A, \theta, a, c)=(0, \pi/4, 1/20, 3)$, we can show that the electromagnetic amplification factor ${\gamma}_{\rm em}(t=200,\rho=1000)/ {\gamma}_{\rm em}(t=0,\rho=1000)$ exceeds 1000 numerically.

Finally we shall give a few comments related to the spacetime structure, the detailed analysis of which will be given in a subsequent paper.
First, it should be noticed that as long as the WWB solution is used, the space-time structure will be expected to be regular because the behavior of the C-energy is the same as the previous vacuum solutions \cite{Mishima:2017zpn}. 
Next, for the deficit angle $\Delta \phi$ of the solutions presented above, we can know that $\Delta \phi \approx O(1)$ by using the formula 
$\Delta \phi = 2\pi (1-e^{-(c/2a)^2})$~\cite{Mishima:2017zpn}. 
For further discussion, let us introduce an ordinary units (e.g., SI base units). Then, by using a dimensional analysis the deficit angle is roughly given, as follows, 
\begin{eqnarray}
\Delta \phi\approx k\frac{G}{c^4}\mu, 
\end{eqnarray}
where $G$ and $c$ represent Newton constant and the speed of light, respectively. 
The above factor $k$ is a some numerical constant of $O(1)$ and $\mu$ is linear energy density.
If the order of the deficit angle is $O(1)$, the corresponding linear energy density $\mu$ can be roughly regarded as $c^4/2G$.
When some mass scale $M$ (e.g. solar mass $M_{\odot}$) is introduced here, $\mu$ is represented as follows, 
\begin{eqnarray}
\mu \approx \frac{c^4}{2G}=\frac{Mc^2}{r_g}, 
\label{eq:linedensity} 
\end{eqnarray}
where $r_g = {2MG}/{c^2}$ (i.e., gravitational radius). 
This linear energy density may be considered as that of an array of an infinite number of black holes having mass $M$ aligned equally at distance $r_g$. 
According to the so-called hoop conjecture~\cite{Thorne1972} applied to elongated and finite objects (for ${\rm LINEAR\ ENERGY\ DENSITY} \gtrsim {Mc^2}/{r_g}$), it should be noticed that a black hole may be formed in the case of linear energy density larger enough than~(\ref{eq:linedensity}). 
Hence, when we consider the conversions of more realistic gravitational and electromagnetic waves which enemate from elongated finite energy distributions, the black hole formation may prevent the spread of the waves. 
As a result, the corresponding conversions may be suppressed in comparison with the cylindrically symmetric solutions of the same linear energy density because of the confinement of the conversion region inside the black hole. 

\acknowledgements
TM was supported by the Grant-in-Aid for Scientific Research (C) [JSPS KAKENHI Grant Number~20K03977], and ST was supported by the Grant-in-Aid for Scientific Research (C) [JSPS KAKENHI Grant Number~21K03560] from the Japan Society for the Promotion of Science.


\end{document}